\documentclass[]{spie}  

\usepackage{amsmath,amsfonts,amssymb}
\usepackage{graphicx}
\usepackage[colorlinks=true, allcolors=blue]{hyperref}
\usepackage{multirow}

\title{Science operation, data handling, and ground support system of the SOLAR-C mission}

\author[a]{Shin Toriumi}
\author[a]{Mari Nishiyama}
\author[a]{Keiichi Matsuzaki}
\author[a,b]{Toshifumi Shimizu}
\author[a]{Hideki Kato}
\author[a]{Jun Matsumoto}
\author[a]{Mizuho Uchiyama}
\author[a]{Kota Miyoshi}
\author[a]{Daiki Yamasaki}
\author[c]{Hirohisa Hara}
\author[c]{Yukio Katsukawa}
\author[c]{Masahito Kubo}
\author[c]{Ryohko Ishikawa}
\author[c]{Takenori J. Okamoto}
\author[b]{Shinsuke Imada}
\author[d]{Satoshi Masuda}
\author[e]{Kyoko Watanabe}
\author[f]{Yusuke Iida}
\author[g]{Ayumi Asai}
\affil[a]{Institute of Space and Astronautical Science, Japan Aerospace Exploration Agency,
3-1-1 Yoshinodai, Chuo-ku, Sagamihara, Kanagawa 229-8510, Japan}
\affil[b]{Department of Earth and Planetary Science, Graduate School of Science, The University of Tokyo, 7-3-1 Hongo, Bunkyo-ku, Tokyo 113-0033, Japan}
\affil[c]{National Astronomical Observatory of Japan, Mitaka, Tokyo 181-8588, Japan}
\affil[d]{Institute for Space-Earth Environmental Research, Nagoya University,
Furo-cho, Chikusa-ku, Nagoya 464-8601, Japan}
\affil[e]{National Defense Academy, 1-10-20 Hashirimizu, Yokosuka, Kanagawa 239-8686, Japan}
\affil[f]{Niigata University, 8050 Ikarashi 2-no-cho, Nishi-ku, Niigata 950-2181, Japan}
\affil[g]{Astronomical Observatory, Graduate School of Science, Kyoto University, Kitashirakawa-oiwake-cho, Sakyo-ku, Kyoto 606-8502, Japan}

\authorinfo{Further author information: (Send correspondence to S.T.)\\S.T.: E-mail: toriumi.shin@jaxa.jp, Telephone: +81 (70) 1170 2784}

\begin{document} 
\maketitle

\begin{abstract}
SOLAR-C is an international solar-observing satellite mission led by Japan Aerospace Exploration Agency (JAXA). It aims to elucidate mass and energy transport in solar atmospheres through extreme ultraviolet (EUV) spectroscopy. The mission carries the EUV High-throughput Spectroscopic Telescope (EUVST) and the Solar Spectral Irradiance Monitor (SoSpIM), enabling comprehensive observations across a wide temperature range ($10^4$ K to $10^7$ K) with minimum temperature gaps and high spatial and temporal resolution. To achieve its science objectives regarding atmospheric heating and solar flare eruptions, SOLAR-C implements a flexible and responsive operational procedure and a data processing system,
while building on the heritage of the Hinode and IRIS satellites. The Chief Observer creates observation timelines that include core observation plans and approved proposed plans while taking into account the solar activity levels. A SpaceWire communication architecture is employed onboard the spacecraft, and a method is implemented in which the mission instrument temporarily acts as the network master during data transfer to support the high data rate requirements. Telemetry is downlinked at ground stations worldwide and gathered at the Institute of Space and Astronautical Science, JAXA. The EUVST data are calibrated at the SOLAR-C Science Center at Nagoya University, while the SoSpIM data are calibrated at the Processing and Archiving Facility before being integrated into the science data products. The data will be made publicly available immediately. The integrated operational scheme for this mission is expected to advance our understanding of solar atmospheric heating and flare processes.
\end{abstract}


\section{INTRODUCTION}
\label{sec:intro}  

The Sun's outer atmosphere is heated by magnetic fields, forming high-temperature structures ranging from the chromosphere, which reaches temperatures of a few $10^4$ K, to the corona, where temperatures exceed 1 MK.\cite{pub:klimchuk2006,pub:reale2014} Some of this coronal plasma is accelerated and flows out into the interplanetary space as the solar wind, creating the heliosphere, the sphere of influence of solar plasma. Solar flares, the largest sudden explosive phenomena in the solar system, also occur in the corona.\cite{pub:shibata2011,pub:toriumi2019} The accompanying increase in electromagnetic radiation across all wavelengths, coronal mass ejections, and solar energetic particles cause significant disturbances in the magnetic and atmospheric environments around the Earth, which are often referred to as space weather phenomena. Understanding the origin of such high-temperature plasma and high-energy phenomena requires clarifying the processes of mass and energy transport within the solar atmosphere. This is the primary objective of the SOLAR-C mission.\cite{pub:shimizu2019,pub:shimizu2021}

   \begin{figure} [tb]
   \begin{center}
   \begin{tabular}{c} 
   \includegraphics[width=0.7\linewidth]{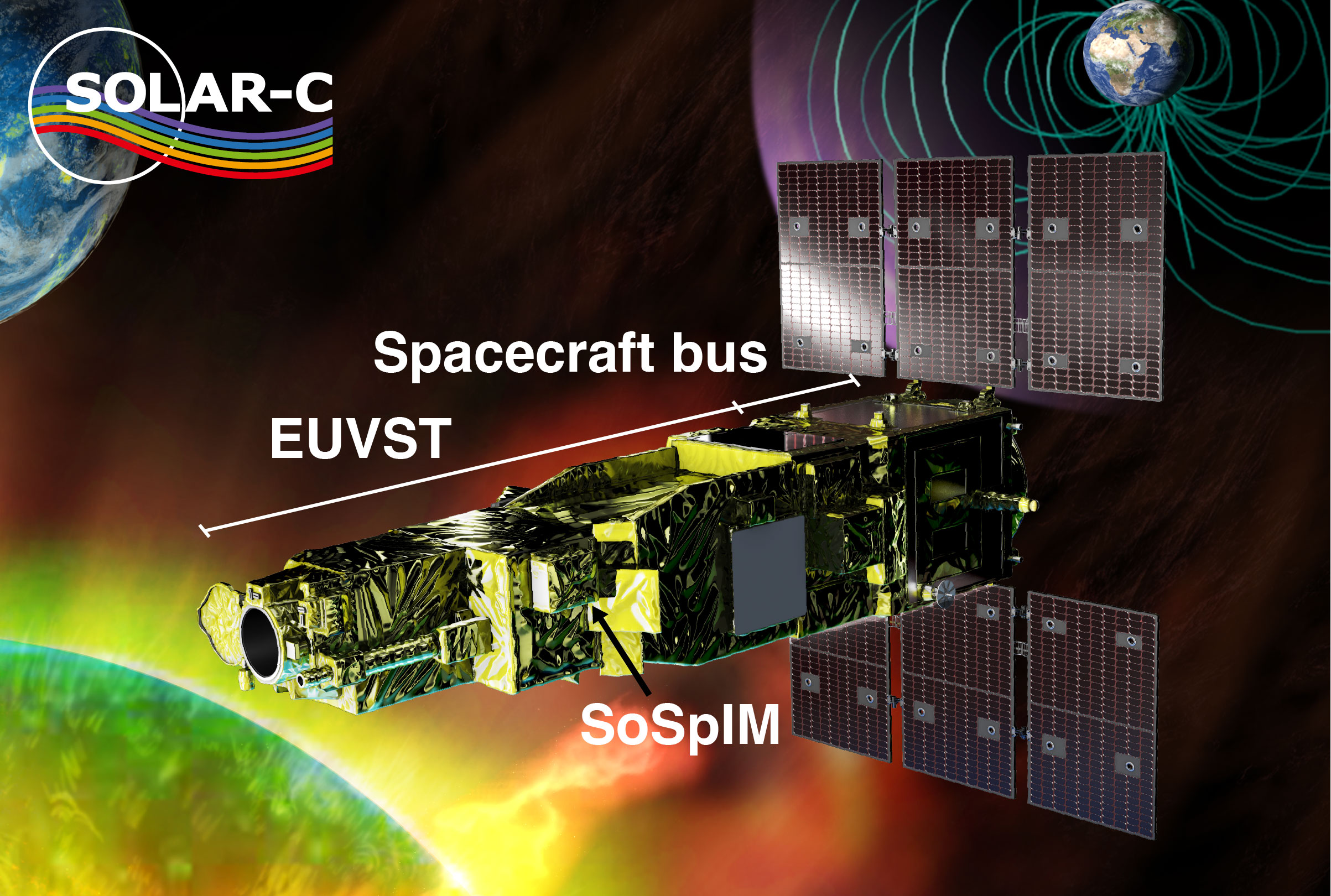}
   \end{tabular}
   \end{center}
   \caption[example] 
   { \label{fig:solar-c} 
Exterior view of the SOLAR-C satellite. It consists of an approximately 1‑m cubic spacecraft bus, on top of which the EUVST instrument with a length of about 3.7 m is mounted. The SoSpIM is mounted on the side of the EUVST telescope structure.}
   \end{figure} 

SOLAR-C (Fig.~\ref{fig:solar-c}) is an international solar-observing mission led by Japan Aerospace Exploration Agency (JAXA) in collaboration with the National Aeronautics and Space Administration (NASA), European Space Agency (ESA), and other European partners. Its primary payload is the Extreme UltraViolet High-throughput Spectroscopic Telescope (EUVST). The EUV wavelength band contains emission lines corresponding to a wide range of temperatures, and by leveraging its broad wavelength coverage, EUVST will perform high-sensitivity, high-spatial-resolution, and high-temporal-resolution spectroscopic observations that cover a nearly continuous temperature range from the chromosphere (20,000 K) to the corona (exceeding 1 MK) and solar flares (approximately 15 MK). This will enable unprecedentedly comprehensive spectroscopic observations of the Sun’s EUV radiation. Meanwhile, the Solar Spectral Irradiance Monitor (SoSpIM) aboard SOLAR-C will perform photometric observations of UV radiation from the entire solar disk, capturing temporal variations in solar brightness with high precision and high temporal resolution in two wavelength bands: the H Ly$\alpha$ 1216 {\AA} line and the Fe IX 171 {\AA} line. These observations will contribute to elucidating the effects of EUV radiation on the Earth’s atmosphere and will also play a crucial role in calibrating the EUVST’s radiometry. The key specifications of EUVST and SoSpIM are summarized in Tab.~\ref{tab:payload}. The integrated and efficient operation of these two instruments is essential for achieving SOLAR-C’s scientific objectives.

\begin{table}[ht]
\caption{Science payload of the SOLAR-C mission.} 
\label{tab:payload}
\begin{center}       
\begin{tabular}{|l|l|l|l|} 
\hline
\rule[-1ex]{0pt}{3.5ex}  \multirow{2}{*}{} & \multicolumn{2}{|l|}{EUVST} & \multirow{2}{*}{SoSpIM} \\
\cline{2-3}
\rule[-1ex]{0pt}{3.5ex}  & Spectrograph & Slit-jaw imager &  \\
\hline
\rule[-1ex]{0pt}{3.5ex}  Spatial resolution &  $0.4''$ & $0.5''$ & N/A \\
\hline
\rule[-1ex]{0pt}{3.5ex}  Field of view & $280''\times 280''$ & $300''\times 300''$ & full-Sun \\
\hline
\rule[-1ex]{0pt}{3.5ex}  Temporal resolution & 0.5 s & 1 s & 100 ms \\
\hline 
\rule[-1ex]{0pt}{3.5ex}  Wavelength & {\raggedright\begin{tabular}{l}17--21.5 nm \\ 46--122 nm\end{tabular}}& \begin{tabular}{l}279.6 nm \\ 283.3 nm \\ 285.2 nm\end{tabular} & \begin{tabular}{l}17.0--21.5 nm (17.1 nm) \\ 111.5--127.5 nm (121.6 nm)\end{tabular} \\
\hline 
\rule[-1ex]{0pt}{3.5ex}  Temperature & 20,000--15 MK & 5000--10,000 K & \begin{tabular}{l}1 MK \\ 30,000 K\end{tabular} \\
\hline 
\end{tabular}
\end{center}
\end{table}

In this paper, Sec.~\ref{sec:2} outlines the concept and science operations of the SOLAR-C mission. Sec.~\ref{sec:3} then describes the observation planning and data handling system, while Sec.~\ref{sec:4} presents the data processing, including the ground support system. Finally, Sec.~\ref{sec:5} summarizes the paper.

\section{MISSION CONCEPT AND SCIENCE OPERATION}\label{sec:2}

\subsection{Science Objectives}

The SOLAR-C mission is designed to understand how the plasma universe is created and evolves, and how the Sun influences the Earth and planets in the solar system. This will be achieved by fulfilling the scientific objectives shown in Tab.~\ref{tab:objectives}. The science operation of SOLAR-C is conducted to maximize the science output and achieve the mission's science objectives by properly operating the spacecraft and mission instruments and acquiring the observation data.

\begin{table}[ht]
\caption{Science objectives of the SOLAR-C mission.} 
\label{tab:objectives}
\begin{center}       
\begin{tabular}{|l|p{14cm}|} 
\hline
\rule[-1ex]{0pt}{3.5ex} No. & Objective/subobjective \\
\hline
\rule[-1ex]{0pt}{3.5ex}  I & Understand How Fundamental Processes Lead to the Formation of the Solar Atmosphere and the Solar Wind \\
\hline
\rule[-1ex]{0pt}{3.5ex}  I-1  & Quantify the Contribution of Nanoflares to Coronal Heating \\
\hline
\rule[-1ex]{0pt}{3.5ex}  I-2  & Quantify the Contribution of Wave Dissipation to Coronal Heating \\
\hline
\rule[-1ex]{0pt}{3.5ex}  I-3  & Understand the Formation Mechanism of Spicules and Quantify Their Contribution to Coronal Heating \\
\hline
\rule[-1ex]{0pt}{3.5ex}  I-4  & Understand the Source Regions and the Acceleration Mechanism of the Solar Wind \\
\hline
\rule[-1ex]{0pt}{3.5ex}  II & Understand How the Solar Atmosphere Becomes Unstable, Releasing the Energy that Drives Solar Flares and Eruptions \\
\hline
\rule[-1ex]{0pt}{3.5ex}  II-1 & Understand the Fast Magnetic Reconnection Process \\
\hline
\rule[-1ex]{0pt}{3.5ex}  II-2  & Identify the Signatures of Global Energy Buildup and the Local Triggering of the Flare and Eruption \\
\hline 
\end{tabular}
\end{center}
\end{table}

\subsection{Operation concept}

To achieve these science objectives, a well-planned and flexible observation strategy is required.
The science operation concept of SOLAR-C incorporates new operational elements and system capabilities to enable more flexible and responsive observations, while utilizing the heritage of Hinode (SOLAR-B)\cite{pub:kosugi2007} and the Interface Region Imaging Spectrograph (IRIS)\cite{pub:depontieu2014}, which have provided stable observations for over a decade, and other preceding solar-observing satellites. This subsection describes the operation concept during the normal observation operation (nominal operation) phase, which spans two years following a four-month commissioning phase.

The EUVST Core Team is responsible for the science operation. The board of the International Science Working Group (SWG) provides overall guidance by leading regular reviews of operations and science activities, and the SWG serves as an interface between the mission team and the broader scientific community.
Prior to launch, the EUVST Core Team will develop core observation plans and core plans (coordinated), aimed at achieving the mission requirements and scientific objectives in Tab.~\ref{tab:objectives}. As part of the report by the Next Generation Solar Physics Mission (NGSPM)'s Science Objectives Team (SOT),\footnote{\url{https://hinode.nao.ac.jp/SOLAR-C/SOLAR-C/Documents/NGSPM_report_170731.pdf}} coordinated observations with NASA's Multi-Slit Solar Explorer (MUSE)\cite{pub:depontieu2019}, which performs high-cadence spectroscopic and imaging observations of the solar corona, have been proposed. Consequently, the core plans (coordinated) will be designed to accommodate coordination campaigns with MUSE. Additionally, SOLAR-C will widely accept observation proposals from researchers worldwide. The Science Schedule Coordinators (SSCs) under the SWG receive and review the proposed plans and schedule the approved plans while considering a good balance with the core plans.

The operational structure differs from that of Hinode in part. On Hinode, three Chief Observers, one for each of its three telescopes, and one Chief Planner, who oversees them, hold discussions to formulate observation plans. In contrast, SOLAR-C has only one telescope, so the process of formulating observation plans will be simpler. There will be one Chief Observer, supported by the Chief Planner.

The Chief Observer will create daily observation timelines. Since the EUVST performs targeted observations with a finite field of view, target regions on the solar disk must be selected. In contrast, the SoSpIM continuously monitors the full solar disk and does not require target selection. Accordingly, the Chief Observer selects target regions and creates observation timelines for EUVST, while only minimal planning is required for SoSpIM. These observation timelines, along with commands for the spacecraft bus, will be uplinked to the spacecraft from the Institute of Space and Astronautical Science (ISAS), JAXA in Sagamihara, Japan. Observation data telemetry will be downlinked via ground stations around the world and gathered at ISAS. Because the occurrence of solar flares is difficult to predict, the team will monitor solar activity to respond rapidly to changes in activity.

\section{SCIENCE PLANNING SYSTEM AND DATA HANDLING}\label{sec:3}

\subsection{Observation Timeline}

   \begin{figure} [tb]
   \begin{center}
   \begin{tabular}{c} 
   \includegraphics[width=0.9\linewidth]{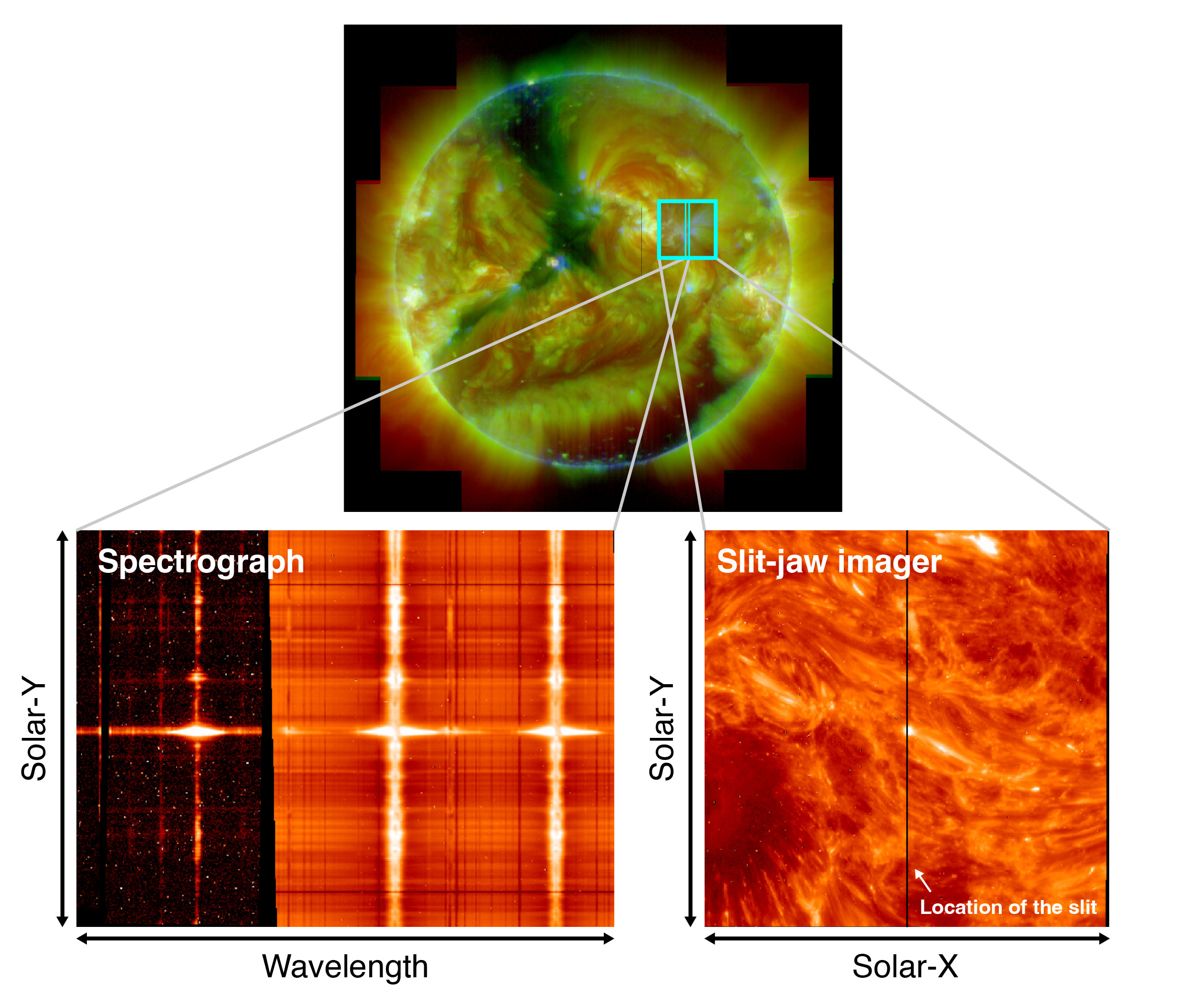}
   \end{tabular}
   \end{center}
   \caption[example] 
   { \label{fig:observation} 
Relationship between the spectrograph and the slit-jaw imager (SJI). When the EUVST slit is placed on a target on the solar disk, the EUV radiation entering the slit is dispersed and captured as spectrograph data by the cameras. The surrounding light that does not pass through the slit is instead captured by the SJI as a two-dimensional context map. This context information is used to determine the slit position on the solar disk by comparison with observations from ground-based observatories and other space telescopes. This demonstration figure is created using data from Hinode (top) and IRIS (bottom).}
   \end{figure} 

To support the operation concept outlined in Sec.~\ref{sec:2}, a dedicated science planning and data-handling scheme has been developed. The creation of observation timelines will utilize the planning framework established for Hinode and IRIS. As a baseline, on weekdays, the Chief Observer creates a one-day observation timeline for the EUVST each day.
The EUVST consists of a spectrograph (three long-wavelength (LW) cameras and one short-wavelength camera (SWC)) and a slit-jaw imager (SJI) (Fig.~\ref{fig:observation}). The Chief Observer monitors solar activity and, taking into account the activity levels, plans and executes predefined core plans and approved proposed plans using these instruments.

The Chief Observer sets both spacecraft parameters (pointing and roll angles) and EUVST instrument parameters.
Since the EUVST is fixed to the spacecraft bus, the EUVST pointing, i.e., the location on the solar disk to be targeted, is directly determined by the spacecraft pointing.
While the pointing was determined by the Chief Planner on Hinode, this task is assigned to the Chief Observer on SOLAR-C due to differences in instrument configuration: Hinode carried multiple telescopes, whereas SOLAR-C carries only one.
The pointing is up to $1.4$ solar radii from the disk center. Additionally, the roll angle represents a new operational feature that was not available on Hinode, allowing the orientation of the slit of the spectrograph on the solar disk to be adjusted to an arbitrary angle. The instrument parameters define the selection of spectral lines, exposure time, spatial binning, scan width and step, and scan cadence, which determine the spatial and temporal resolution of the EUVST observations.

The Chief Observer also conducts coordination with ground-based observatories and other space telescopes. It is under consideration that the observation timeline creation tool will be remote-accessible. This feature will allow the Chief Observer to create observation timelines from remote locations. The timelines also include regular calibration campaigns. For example, full-disk mosaics will be taken for cross-calibration between EUVST and SoSpIM. The top panel of Fig.~\ref{fig:observation} shows a sample full-disk mosaic obtained by Hinode.

\subsection{Onboard Data-handling System}

   \begin{figure} [tb]
   \begin{center}
   \begin{tabular}{c} 
   \includegraphics[width=0.9\linewidth]{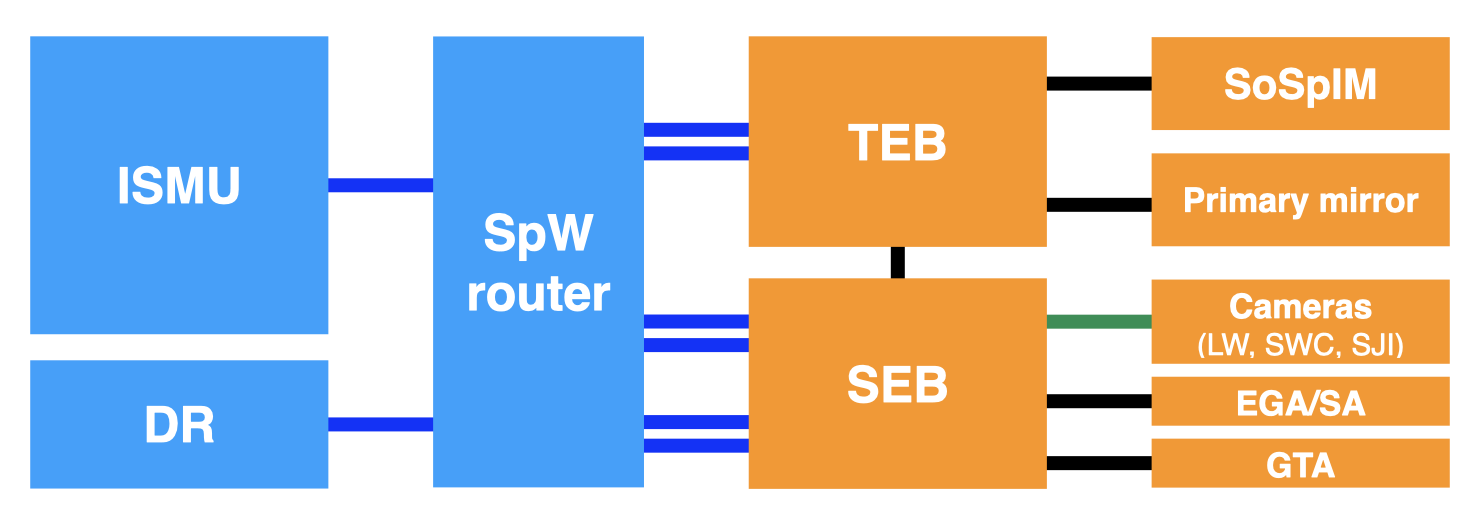}
   \end{tabular}
   \end{center}
   \caption[example] 
   { \label{fig:network} 
Configuration of the onboard SpW data-handling network of SOLAR-C. The network connects the spacecraft bus (ISMU and DR) with the EUVST components (SEB and TEB).
Blue lines represent the SpW network of the spacecraft bus, while lines in other colors indicate signal lines on the EUVST side.}
   \end{figure} 

The SOLAR-C spacecraft bus is equipped with an Integrated Satellite Management Unit (ISMU), which manages the entire satellite. The ISMU acts as a network master, forming a SpaceWire (SpW) data-handling network that includes the mission segment. The overall configuration of the SpW network is illustrated in Fig.~\ref{fig:network}.

The Spectrograph Electronics Box (SEB) and the Telescope Electronics Box (TEB) control the mission payload, the EUVST, and are both connected to the SpW network. The SEB controls the spectrograph subsystem components, operating the cameras (LW, SWC, and SJI), grating assembly (EGA), slit assembly (SA), and guide telescope assembly (GTA). The SEB's internal sequencer and flight software operate all of these peripheral components. The TEB controls the tip-tilt mechanism of the primary mirror assembly.

The SEB acquires image data from each camera, processes them (including image compression), and transmits the generated data to the spacecraft bus data recorder (DR) via the SpW network. Depending on the Chief Observer's settings, the flight software can analyze obtained image data, detect solar flares, and automatically control the exposure time. For image compression, EUVST employs the CCSDS Image Data Compression (both lossy and lossless).

The ISMU maintains time synchronization with the SEB and TEB by broadcasting time indicators over the SpW network. Additionally, the ISMU defines time slots and controls the timing of communication. On many JAXA scientific satellites, the ISMU constantly acts as the network master to send commands and collect telemetry. However, since the transmission rate of observation data on SOLAR-C is high, up to approximately 9 Mbps, a new methodology has been introduced. In this methodology, the SEB acts as the network master during specific time slots to actively send observation data to the DR. The SpW network between the spacecraft bus and SEB/TEB is designed to provide redundancy and enable mitigation against malfunctions.

The TEB receives signals from GTA via the SEB, which indicate the offset error of the solar disk within the EUVST field of view. By using this, the TEB controls the primary mirror with high precision. Additionally, the SoSpIM is connected as a downstream component of the TEB. The observation data from the SoSpIM are collected by the ISMU as part of the TEB telemetry and stored in the DR.

\section{GROUND SUPPORT SYSTEM AND DATA DISTRIBUTION}\label{sec:4}

\subsection{Data downlink}

The SOLAR-C telemetry, including observation and housekeeping data, is downlinked through ground stations around the world. In addition to JAXA's network of domestic and overseas ground stations, SOLAR-C uses high-latitude stations provided by ESA. The telemetry received at each ground station is eventually transmitted to and gathered at ISAS. The requirement for the downlink volume of science data is 72 Gbits per day at maximum.

\subsection{Data Utilization System}

   \begin{figure} [tb]
   \begin{center}
   \begin{tabular}{c} 
   \includegraphics[width=0.7\linewidth]{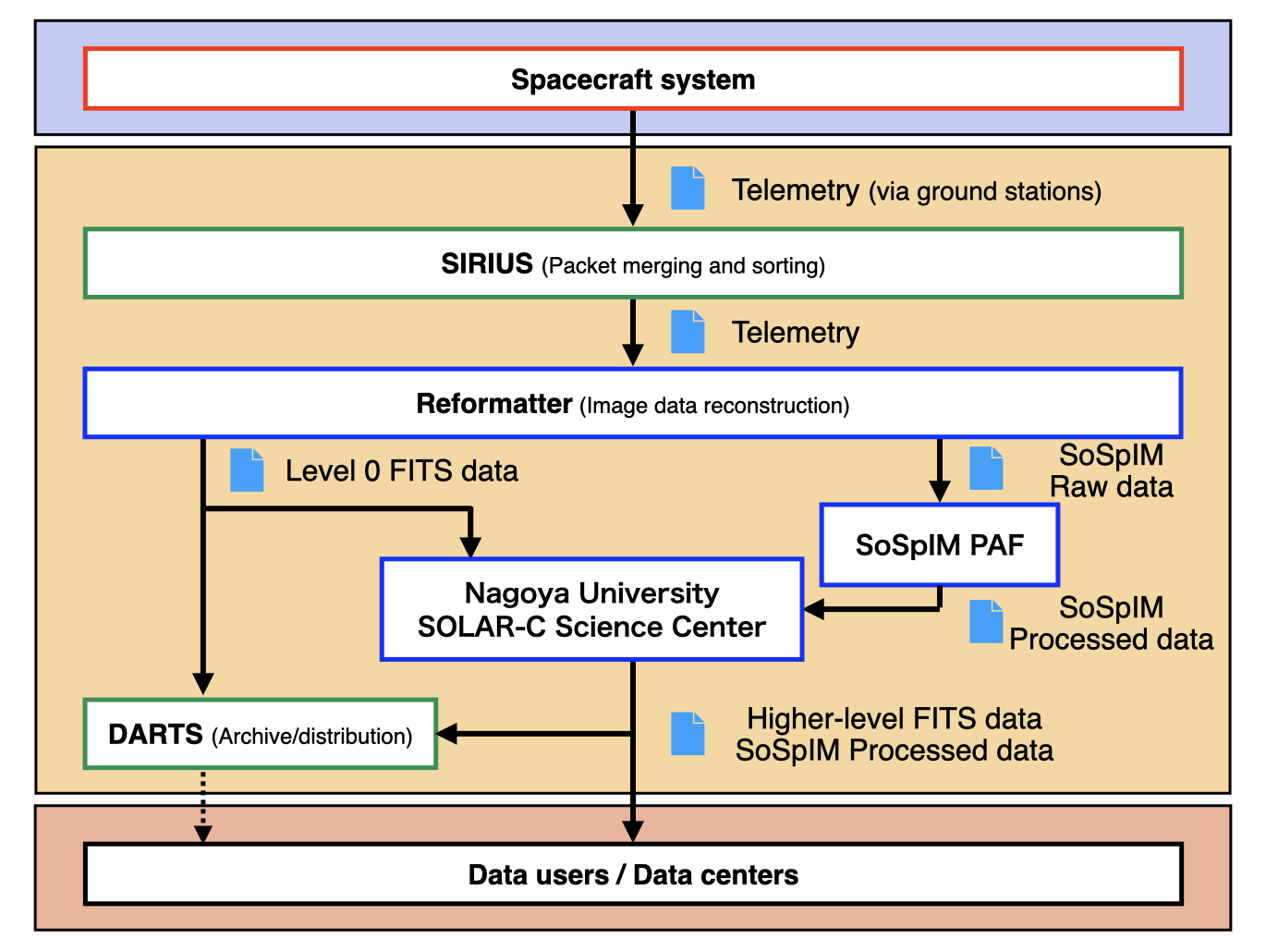}
   \end{tabular}
   \end{center}
   \caption[example] 
   { \label{fig:flow} 
Flowchart of the SOLAR-C data processing. The left branch indicates the EUVST data, while the SoSpIM data flow is shown on the right-hand side.}
   \end{figure} 

An overview of the data processing flow of the SOLAR-C mission is shown in Fig.~\ref{fig:flow}. All telemetry data collected at ISAS are archived after time stamping, removing the duplicates, and sorting them by time on the SIRIUS (Scientific Information Retrieval and Integrated Utilization System) database. For science purposes, EUVST observation data are converted into Level 0 Flexible Image Transport System (FITS) files on the reformatter. These data are then transmitted to the SOLAR-C Science Center at Nagoya University. There, calibration is performed on the Level 0 data, and higher-level data (Level 1 and higher) are generated. The calibration plan is shown in Tab.~\ref{tab:fits}. The generated data and analysis software are made publicly available worldwide from the SOLAR-C Science Center.

\begin{table}[ht]
\caption{Levels of the EUVST data.} 
\label{tab:fits}
\begin{center}       
\begin{tabular}{|l|l|} 
\hline
\rule[-1ex]{0pt}{3.5ex} Level & Calibration \\
\hline
\rule[-1ex]{0pt}{3.5ex}  0 & Reformatted \\
\hline
\rule[-1ex]{0pt}{3.5ex}  1  & Dark field, flat field, radiometry, geometric correction; assembly of rasters and time series \\
\hline
\rule[-1ex]{0pt}{3.5ex}  2  & Wavelength correction and co-alignment \\
\hline
\rule[-1ex]{0pt}{3.5ex}  3  & Spectral width, Doppler shift, and radiance data via, e.g., Gaussian line fitting \\
\hline 
\end{tabular}
\end{center}
\end{table}

From ISAS, the SoSpIM raw data are transmitted to the Processing and Archiving Facility (PAF) located in Switzerland. After calibration processing at this facility, the data are transferred to the SOLAR-C Science Center, where they are made publicly available.

In addition, the EUVST and SoSpIM data are planned to be copied to NASA and ESA data centers. For the purpose of secure data storage, regular copying of the EUVST and SoSpIM data to DARTS (Data ARchives and Transmission System) on ISAS is planned.

The Science Center manages these data by, e.g., providing data search functionality and assigning DOIs (Digital Object Identifiers). Alignment calibration of observational data and numerical modeling of solar atmospheres are also included in the target tasks of the Science Center. By using them, researchers are able to compare multi-wavelength data and interpret the physical meaning of the data.

\subsection{Data policy}

All data will be made publicly available as soon as the necessary calibration is complete.
In addition to these standard data products, the generation and prompt release of quick-look data that may be useful for space weather studies are also under consideration. This open and timely data policy promotes rapid scientific utilization and collaboration within the international scientific community.

\section{Summary and Conclusion}\label{sec:5}

This paper provides an overview of the science operation, data handling, and ground support system of the SOLAR-C mission. SOLAR-C is an international mission that aims to elucidate energy and mass transport processes in the solar atmosphere through EUV spectroscopy. Achieving its science objectives requires highly integrated operational and data processing systems.

The science operation of SOLAR-C introduces new elements that enable flexible and responsive observations while building upon the heritage of Hinode and IRIS.
With an operational structure that includes a Chief Observer, a Chief Planner, and the EUVST core team, daily observation timelines can be created flexibly, enabling rapid responses to changes in solar activity, especially solar flares. Scientific outcomes will be maximized by combining the core plans and proposed plans. The SWG will provide advice as needed to ensure that observations are conducted appropriately.

New methodologies have been introduced in data handling for onboard communication and data transfer to efficiently acquire spectroscopic data with high temporal and spatial resolution. One of the essential features for meeting the high data rate requirements is
the SpW network methodology: while the ISMU collects regular telemetry from the SEB and TEB as the network master, for observation data transfer, the SEB temporarily becomes the master within allocated time slots and actively sends data to the DR.

Key roles of the ground support system include acquiring data through a global network of ground stations and collecting them at ISAS, as well as the data processing and calibration at the SOLAR-C Science Center and PAF. Observation data from EUVST and SoSpIM will be converted into high-level data through processing pipelines, and the generated high-level data will be promptly released from the Science Center. This immediate release policy is expected to promote data utilization and scientific research by researchers worldwide.

As described above, SOLAR-C is designed as a comprehensive system that integrates the spacecraft, observational instruments, science operation, data processing, and data distribution. This mission is expected to significantly advance our understanding of the heating mechanisms in the solar atmosphere and the physical processes underlying flare eruptions, and is anticipated to have a substantial impact on solar physics and related fields.

\acknowledgments 
The authors used an AI-based tool to assist with grammar checking.

\bibliography{report} 
\bibliographystyle{spiebib} 

\end{document}